\documentclass{llncs}
\usepackage{llncsdoc}
\usepackage{amssymb,amsmath}
\usepackage{cite}
\usepackage{graphicx}
\usepackage{subfig}
\usepackage{array}
\usepackage{cases}
\usepackage{url}
\usepackage{algorithm}
\usepackage{algorithmic}
\usepackage{color}

\newtheorem{mydef}{Definition}

\begin{document}


\title{Inference on the Network Evolution in BitTorrent Mainline DHT}

\author{{Liang Wang and Jussi Kangasharju}}

\institute{Department of Computer Science, University of Helsinki, Finland}


\maketitle

\begin{abstract} 
  Network size is a fundamental statistic for a peer-to-peer system
  but is generally considered to contain too little information to be
  useful.  However, most existing work only considers the metric by
  itself and does not explore what features could be extracted from
  this seemingly trivial metric.  In this paper, we show that Fourier
  transform allows us to extract frequency features from such time
  series data, which can further be used to characterize user
  behaviors and detect system anomalies in a peer-to-peer system
  automatically without needing to resort to visual comparisons.  By
  using the proposed algorithm, our system successfully discovers and
  clusters countries of similar user behavior and captures the
  anomalies like Sybil attacks and other real-world events with high
  accuracy.  Our work in this paper highlights the usefulness of more
  advanced time series processing techniques in analyzing network
  measurements.
\end{abstract}

\section{Introduction} \label{sec:introduction}

Measuring the properties of a system can not only report various
system statistics, but can also be used to expose underlying
relationships between general system dynamics and other
characteristics.  This, however, requires that the measurements
capture the essential characteristics or that appropriate processing
is done to the measurements to enable extraction of useful features.
In many existing measurement work, measured data is simply reported
as-is, which makes obtaining deeper insight from the measurements
difficult.  While such ``naive'' results hold some interest in
themselves, more advanced methods yield more insight and enable
automatic comparison of many interesting features.

System measurement has its irreplaceable importance in the
Peer-to-Peer (P2P) research.  The purpose of the measurements is not
merely meant to report the system statistics but also expected to
expose the underlying relations between the system dynamics and other
characteristics.  However, such insight from the statistics can only
be obtained given the data is properly processed, in other words, when
the useful information is extracted.  E.g.  the network size is a
fundamental statistic for a P2P system, and network evolution data
describes how the size changes as a function of time.  However,
network size as a system metric is generally considered containing too
little information to be useful.  On the contrary, the fact is network
evolution data contains much more useful and interesting information
than most people expect.  As we will show in this paper, such
information can help us finding similar use patterns of different
countries and even detecting system anomalies.

In this paper we use Fourier transform to extract representative
features from a time series of measurements of the size of BitTorrent
MLDHT network.  Fourier transform is not sensitive to issues like
temporal alignment of different samples and it provides us with an
easy way of obtaining a ``fingerprint'' of the data sample.  We apply
this technique to measurements in different ways and show how it can
be used to identify many relevant features of the network.  Fourier
transforms are a well-known and widely-used technique in other fields,
but they have not so far been much used in network measurement work.

Specifically, our contributions are as follows:

\begin{enumerate}

\item We use Fourier transforms to extract frequency-based features
  from network evolution data.  The simple algorithm is also
  applicable to other time series data in P2P measurements.

\item We show the frequency-based features are robust to the noisy
  data and can be used to characterize user behaviors and detect
  system anomalies.  Our method indicates the data sets with simple
  structure may contain more information than we have expected.

\item We implement our algorithm and evaluate it on a realistic
  monitoring system.  We report its actual performance along with some
  interesting findings.

\end{enumerate}

The rest of the paper is organized as follows.
Section~\ref{sec:background} gives a brief introduction on Fourier
transform and related work.  Section~\ref{sec:core} presents our
method of extracting frequency-based features and
Section~\ref{sec:exp} evaluates it on the realistic system.  Finally,
Section~\ref{sec:conclusion} concludes the paper.

\section{Background on Fourier Transform}
\label{sec:background}

The original idea of Fourier transform is to decompose a
function\footnote{In this paper, we use function, signal and time
  series interchangeably as long as it does not cause ambiguity in the
  context.} into a summation of a series of sinusoids waves, or in
other words, projecting a function from one function base to another.
Often, a discretized form consisting only of $T$ consecutive sampling
points is used which leads to the discrete Fourier transform.  In
complex form, it can be written as follows:






\begin{align}
  & f[t] = \sum_{k=0}^{T-1} \widehat{f[k]} e^{j 2\pi k t / T} \quad \quad t \in \{0, 1, 2 ... T-1\} \label{eq:d:5} \\
  & \widehat{f[k]} = \frac{1}{T} \sum_{t=0}^{T-1} f[t] e^{-j 2 \pi k t
    / T } \quad k \in \{0, 1, 2 ... T-1\}
\end{align}

where $j$ is the imaginary unit and $\widehat{f[k]}$ denotes the
Fourier coefficient of frequency $k$ in the complex form.  Fourier
transform in the complex plane is much simpler than the trigonometric
form and can be efficiently calculated using \textit{Fast Fourier
  Transform} (FFT).
For a more thorough overview of Fourier transform and related
techniques, please refer
to~\cite{bracewell1980fourier,bloomfield2004fourier}.



Despite of its wide application in many fields, Fourier transforms (or
other similar methods) have not been commonly used in peer-to-peer
research even though a lot of research has focused on collecting
data~\cite{6688697, 6688698, 6688699, 6335811, 5721044, 5482574,
  6335802}.  The same is largely true of many other network
measurement research.  Though in many existing works, the focus is on
different aspects of the system ranging from security and topology to
economic incentives, the general methodology is in many cases similar
in the sense that the goal is to find a pattern by extracting as much
information as possible from time series data.  For this kind of work,
Fourier transforms and other mechanisms, such as wavelets, can provide
interesting insight.
In this paper we show that using Fourier transforms, we are able to
extract meaningful insight from a simple time series of network size
evolution data.  Past works that have looked at this kind of data
typically have been limited to describing quantitative numbers about
the size evolution or usage in different countries.  Our work shows
that we can classify different countries according to their usage
behavior and easily visualize how different countries group together.
Naturally, more sophisticated methods, such as wavelets, could also be
used and would likely yield additional insights; our goal in this
paper is to highlight the benefits of using proper time series
analysis techniques and open the road for further research.


\section{Frequency-Based Feature}
\label{sec:core}

\begin{figure}[!tb]
  \centering
  \includegraphics[width=10cm]{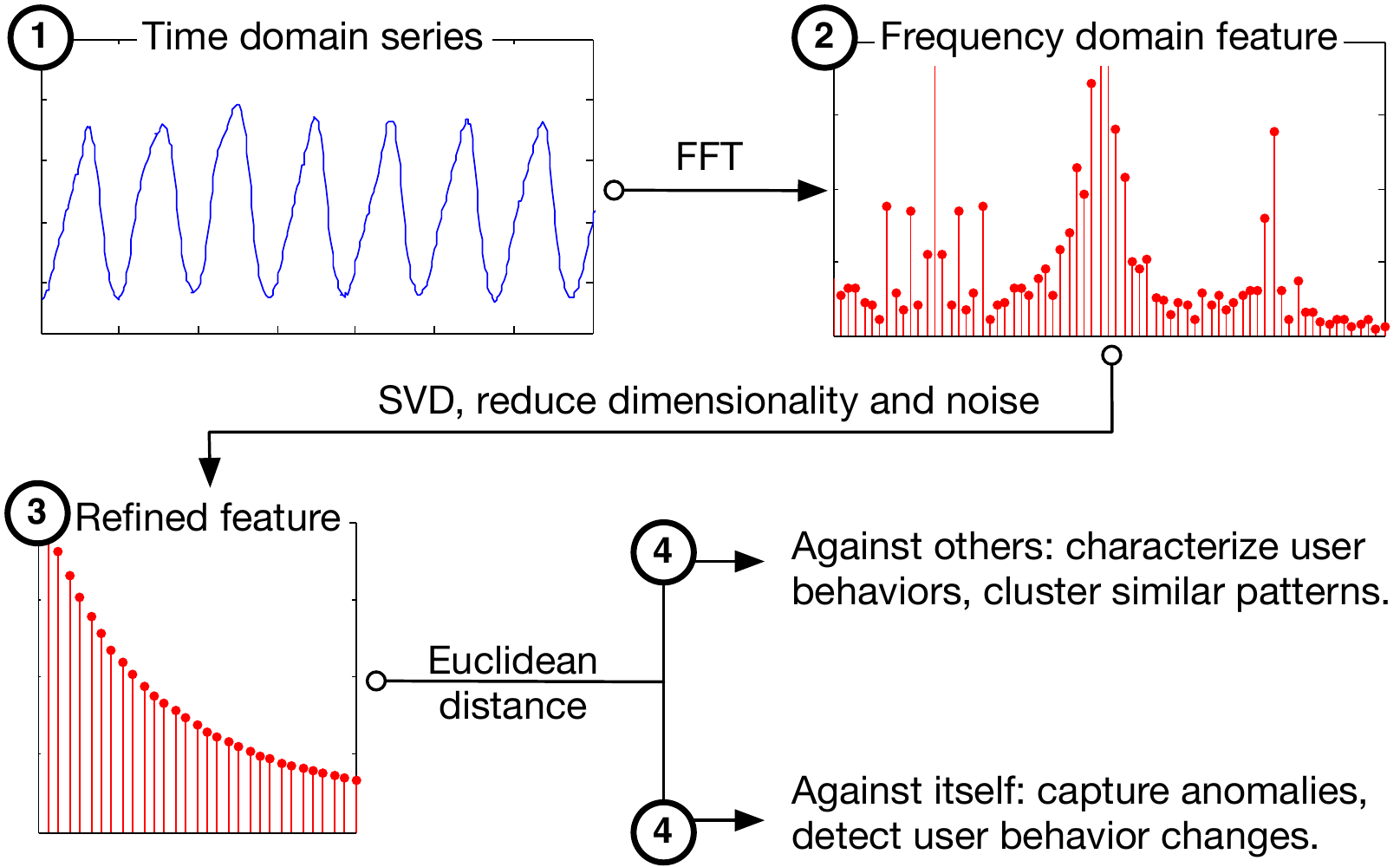}
  \caption{The general workflow from feature extraction to user
    behavior characterization and anomaly detection.}
  \label{fig:arch}
\end{figure}

To compare user behavior and detect behavior changes, as well as to
capture anomalies, we first need to identify a robust feature set.  As
mentioned, we apply Fourier transform to country-level network size
evolution and then generate a frequency-based feature set as its
unique ``fingerprint''.  In this section, we formulate the relevant
definitions and detail the actual steps of feature extraction from
network size evolution.  Then we show how the similarities derived can
be used to discover patterns and detect anomalies.
Fig.~\ref{fig:arch} depicts the general workflow of the proposed
method.  Starting from the raw time series data, i.e., estimated
number of users in a country or the whole system, we perform the
Fourier transform to obtain the frequency domain features.  Using
singular value decomposition, we reduce the dimensionality of the
feature vector to allow for easier comparison of different samples.
Below we will detail these steps.

\subsection{Feature Extraction with FFT}
\label{sec:core:feature}

Our data sets are collected from a monitoring system similar to the
one in~\cite{6688697}, which continuously estimates the network size
of MLDHT by crawling its ID space.  Each crawl generates a full list
of IP addresses in a subspace.  Using the methodology presented
in~\cite{6688697}, we can get a good estimate of the total network
size.  System-level network evolution can be obtained by a series of
such measurements, then further decomposed to country-level (even
city-level) evolution based on their IP blocks.  We use the MaxMind
service\footnote{www.maxmind.com} to perform the mapping of IP
addresses to geographical locations.

\begin{mydef}
  Network evolution of a specific country $c$ is a time series
  $f_c(t)$ which gives the network size of $c$ at time $t$.
\end{mydef}

For simplicity of notation, we drop the subscript $c$ if it causes no
confusion in the context.  Our actual measurements show that $f(t)$ is
practically a stationary function with a period of one week for almost
all the countries, because user behavior is generally different in the
weekends than weekdays, but similar from one week to another.  So in
our calculation, we adopt a period window of seven consecutive days
(irrelevant of the starting point) which consists of $T$ discretized
sampling points $f[0]$, $f[1]$, $f[2]$ ...  $f[T-1]$.  Fourier
transform of this time series gives another complex vector of the same
length which we will use as the starting point of the frequency-based
feature set.

\begin{mydef} 
  Frequency-based feature set $F$ of a specific network evolution
  series $f(t)$ is a real vector where element $k$ represents the
  normalized Fourier coefficient associated with frequency $k$ in
  $f(t)$.

  Specifically,
  \begin{align} & F[k] = \frac{\widehat{f[k]}}{\parallel
      \widehat{f} \parallel} = \frac{1}{T \parallel
      \widehat{f} \parallel } {\left| \sum_{t=0}^{T-1} f[k] e^{-j 2\pi
          k t / T} \right|} \label{eq:f:1}
  \end{align}

  where the normalizer $\parallel \widehat{f} \parallel$ equals

\begin{align}
  & \parallel \widehat{f} \parallel = \frac{1}{T}
  \sqrt{\sum_{k=0}^{T-1} \left| \sum_{t=0}^{T-1} f[k] e^{-j 2\pi k t /
        T} \right|^2 } \label{eq:f:2}
\end{align}

\end{mydef}

$\parallel \cdot \parallel$ above denotes Euclidean norm.
E.q.(\ref{eq:f:1}) is simply the Fourier coefficient of frequency $k$
normalized by $\parallel \widehat{f} \parallel$.  E.q.(\ref{eq:f:2})
calculates the length of the raw coefficient vector in $T$-dimensional
space which is used as an normalizer in (\ref{eq:f:1}).  The purpose
of normalization is to filter out the effect from the absolute network
size but focusing on the pure pattern of a curve.  Essentially, we are
using a normalized vector of Fourier coefficients as the feature set,
where each element represents the amplitude of its associated
frequency component in the original signal $f(t)$.

High dimensional features are usually noisy, redundant and expensive
to compare.  Common technique for dimensionality reduction is
\textit{Singular Value Decomposition} (SVD).  Fig.~\ref{fig:pca:1}
shows the semi-log plot of the singular values in a feature set.  We
can see the first two components dominate the system dynamics.  More
precisely, the first principal component itself captures over 77\% of
the energy.  In practice, we only keep the first 40 principal
components (denoted by $u$) in order to retain 99.20\% of the system
energy.  The refined feature set can be calculated with $u$ as
follows, and $k = 40$ in our case.

\begin{align}
  & F^* = [ u_0, u_1, u_2 ... u_{k-1} ]' F \label{eq:f:3}
\end{align}

The actual computation can be done efficiently since FFT and SVD are
standard routines in scientific computing libraries.  Once the feature
set is successfully extracted, we can use it to characterize user
behavior and further look for similar patterns with properly defined
similarity metric as follows.

\begin{mydef}
  Similarity $S_{a,b}$ of two feature sets $F^*_a$ and $F^*_b$ is
  defined as their Euclidean distance in $k$-dimensional space.
  \begin{align}
    & S_{a,b} = \parallel F^*_a - F^*_b \parallel \label{eq:f:4}
  \end{align}
\end{mydef}

The code in Algorithm~\ref{alg:feature} shows how the features are
actually extracted and pairwise similarity is calculated.  The key
input $M$ is a matrix where each column vector represents a country's
network evolution of a 7-day window.  The column vector contains 2048
discretized sampling points.  $k$ is the number of principal
components we want to keep in the final feature set.  The rest of the
code is easy to understand.  Lines 4 - 8 iterate all the countries,
where line 5 calculates the Fourier coefficients using FFT, and line 6
normalizes it to a real unit vector.  Note that in our case
\textit{abs} function translates each complex element in $z$ into
corresponding amplitude.  Lines 9 and 10 reduce the feature set
dimensionality by taking the first $k$ principal components.  Function
\textit{pdist} in line 11 calculates the pairwise distance for a given
point set.  Eventually, each column in the output matrix $F$
represents a $k$-dimensional feature set of a country, and $S$
contains their pairwise similarity metrics.

\begin{algorithm}[!tb]
  \caption{Feature extraction and pairwise similarity}
  \label{alg:feature}
  \begin{algorithmic}[1]
    \STATE {\textbf{Input:} Network evolution matrix $M$, $k$ first
      components} \STATE {\textbf{Output:} Feature matrix $F$,
      similarity matrix $S$} \STATE {[$h$, $w$] = size($M$)}

    \FOR { $i$ = $1$ : $w$ } \STATE{$z$ = fft($M(::,i)$)} \STATE{$z$ =
      abs($z$) / sqrt($z' * z$)} \STATE{$F$ = [$F$ $z$]}
    \ENDFOR

    \STATE {[ $u$, $s$, $v$ ] = svd($F$)} \STATE {$F$ = $u( : , 1 : k
      )' * F$} \STATE {$S$ = pdist($F'$, 'euclidean')}
  \end{algorithmic}
\end{algorithm}

\subsection{Anomaly Detection Based on Fluctuation}
\label{sec:core:anomaly}

\begin{figure}[!htb]
  \centering \subfloat[Semi-log plot of the principal components in
  the feature set. The first two dominate the
  dynamics.]{\label{fig:pca:1}\includegraphics[width=5.2cm]{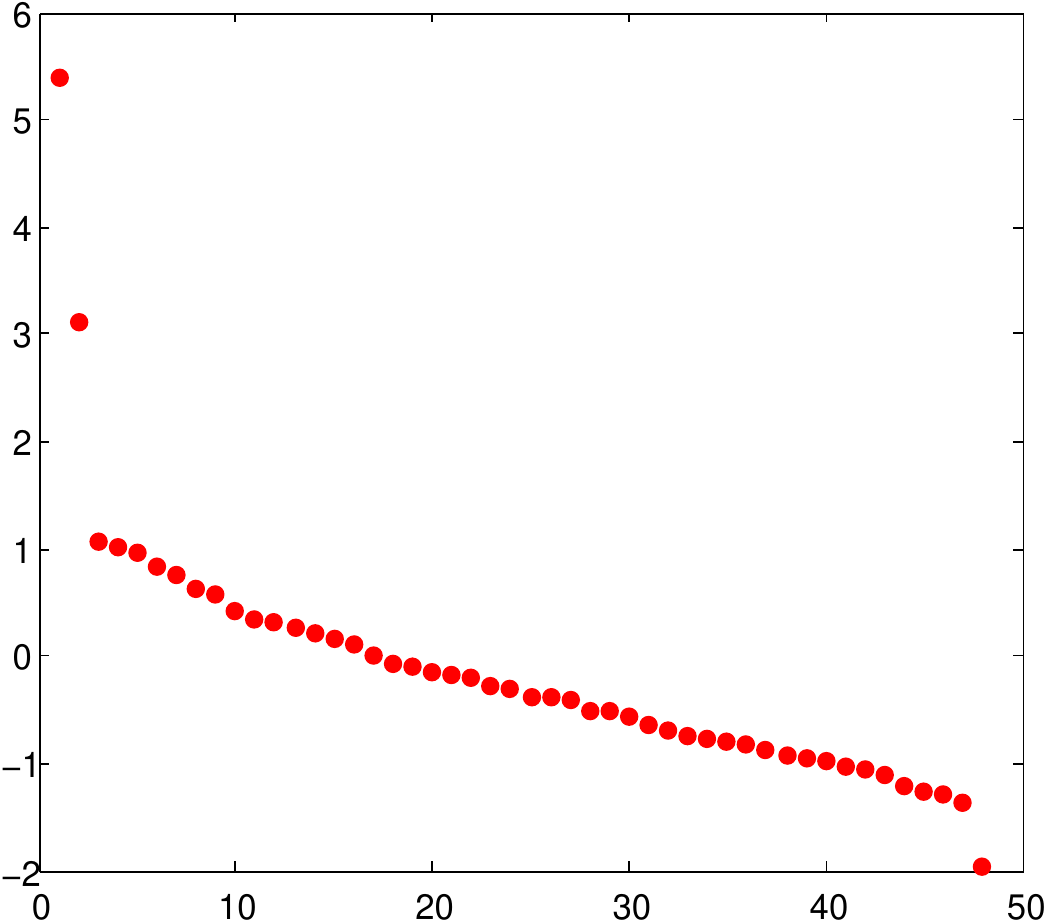}}
  \quad \quad \quad \quad
  \subfloat[Feature fluctuation due to the Sybil-attack, and
  feature distribution during normal
  operations.]{\label{fig:pca:2}\includegraphics[width=5.5cm]{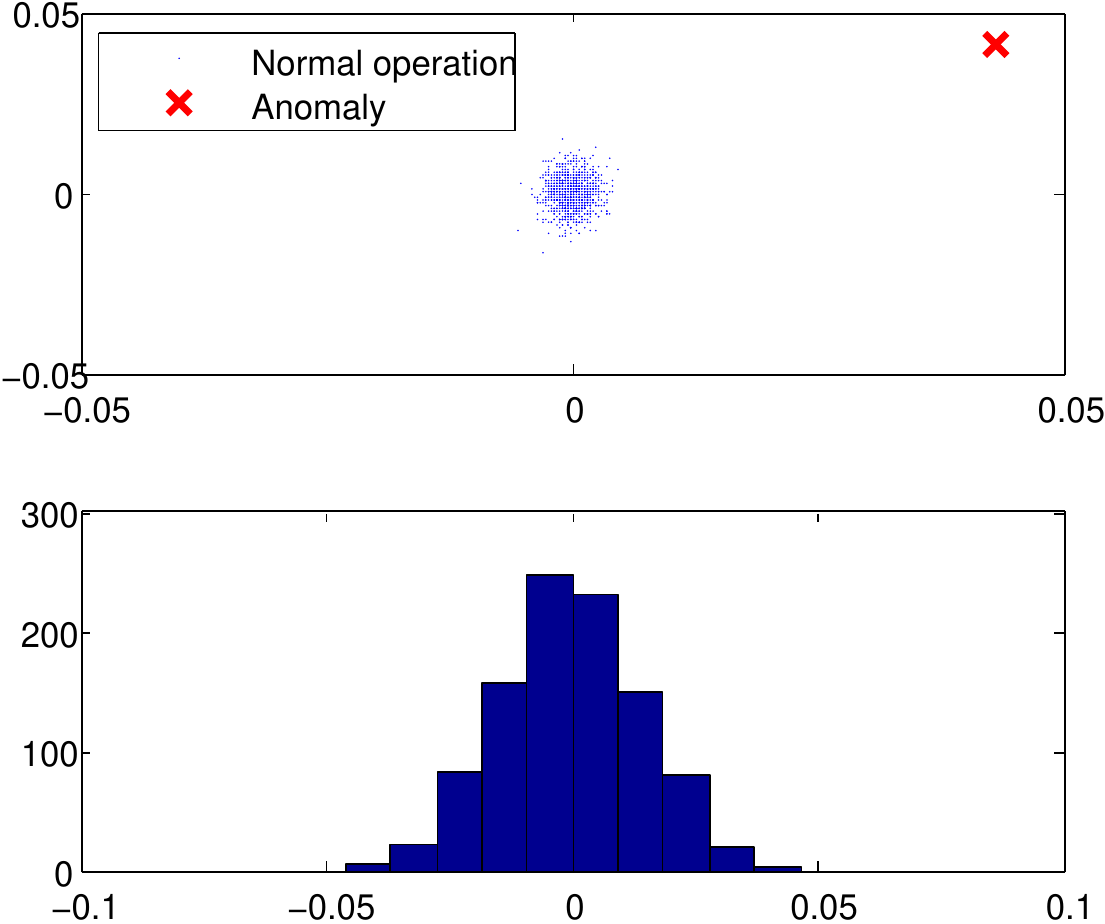}}
  \caption{Feature set is dominated by only few components, and is
    rather stable during the system normal operations. }
  \label{fig:pca}
\end{figure}

In a broad sense, a system anomaly indicates a significant change in a
certain system metric which is usually reflected as fluctuation or
drift in the corresponding time series.  Such anomalies might be due
to service breakdowns, system attacks, or drastic changes in use
pattern.  Detecting such anomalies with satisfying sensitivity and
fall-out rate is generally difficult, especially with noisy data.

\begin{figure*}[!tb]
  \centering \subfloat[Similarity is extracted from simple network
  evolution
  series.]{\label{fig:cluster:1}\includegraphics[width=11cm]{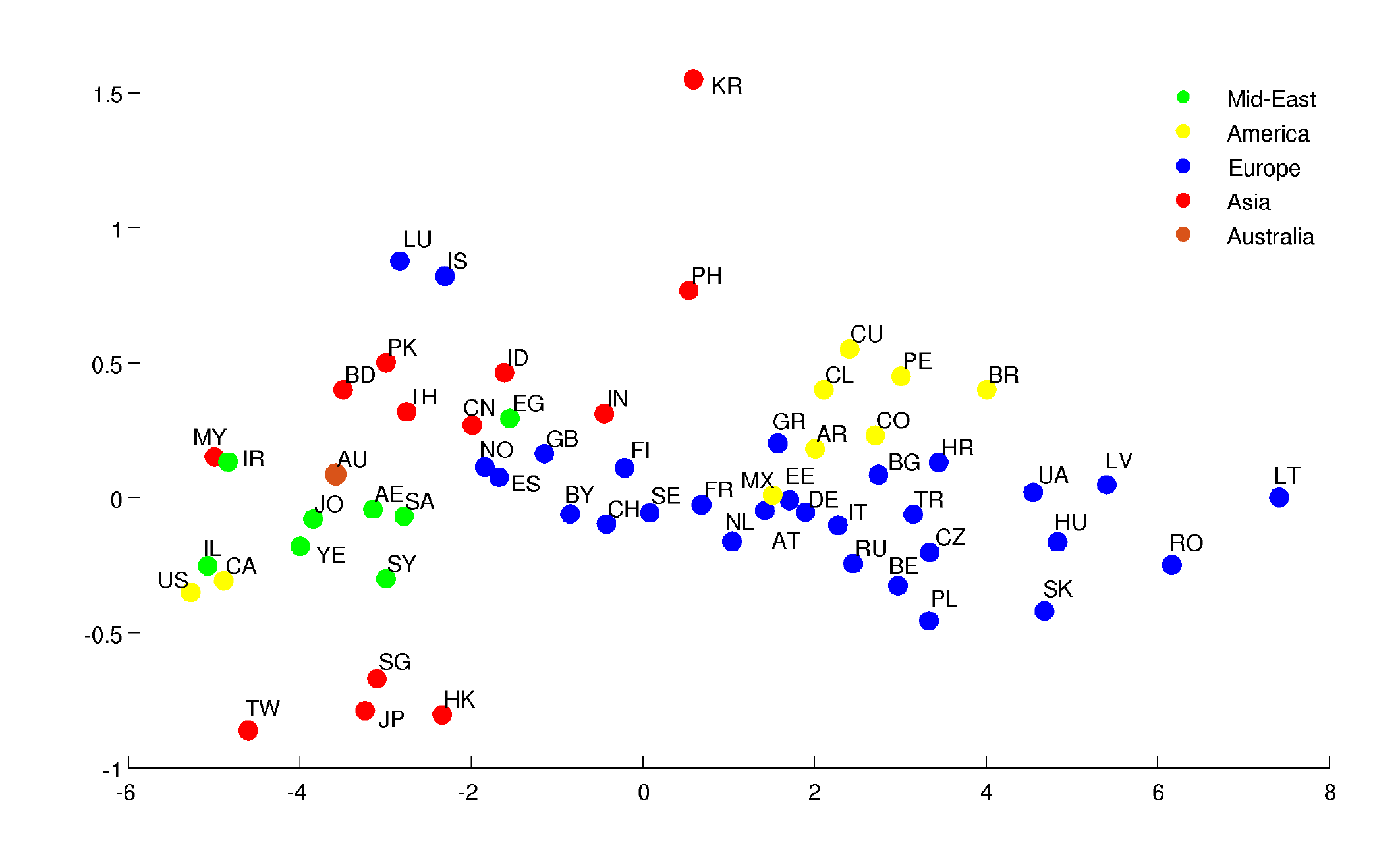}}
  \quad \subfloat[JP, TW and
  KR]{\label{fig:cluster:2}\includegraphics[width=5.5cm]{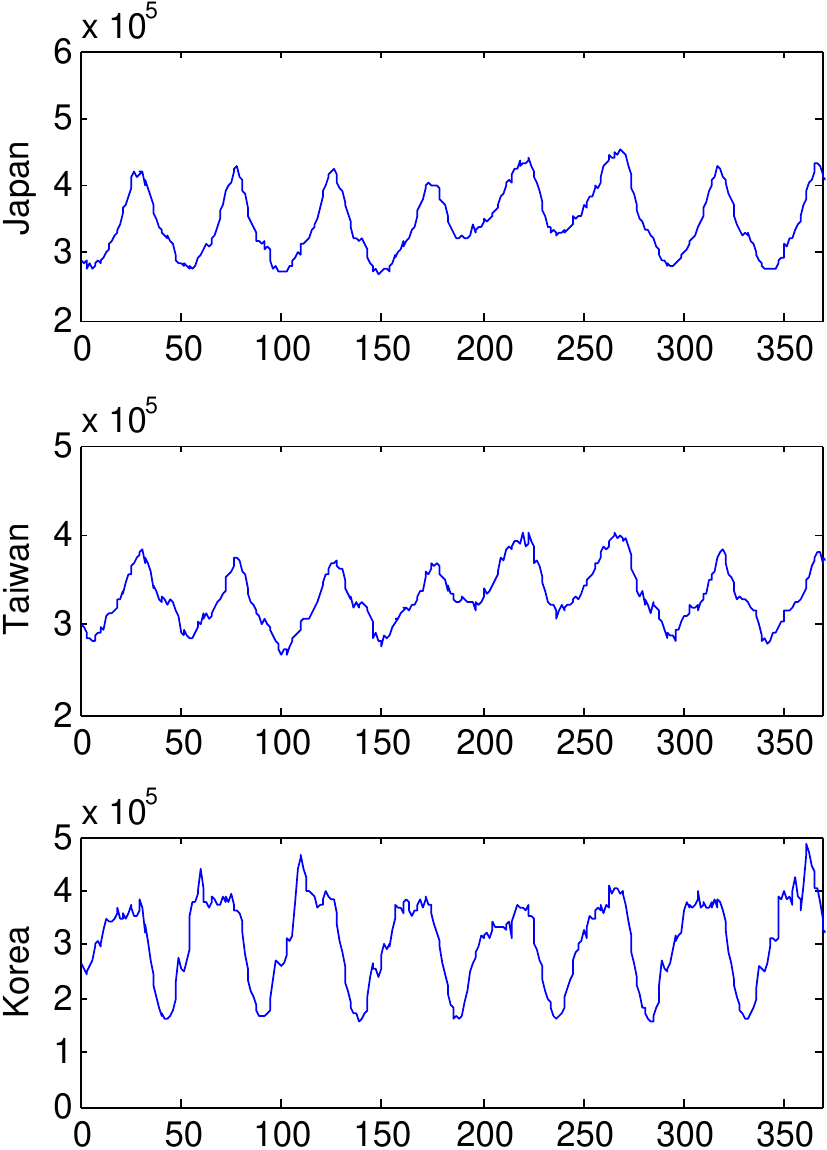}}
  \quad \subfloat[US, CA and
  RU]{\label{fig:cluster:3}\includegraphics[width=5.5cm]{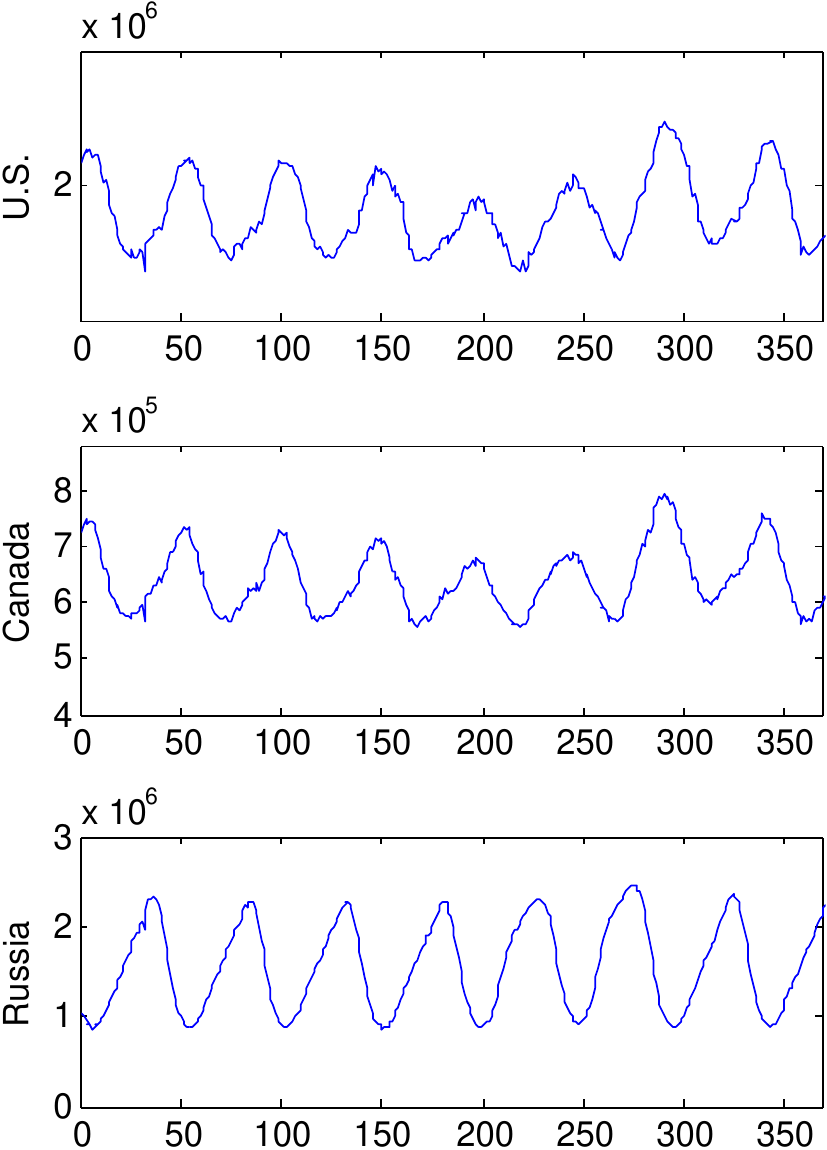}}
  \caption{Clustering of countries based on the similar use patterns,
    and the original network evolution of some selected countries.}
  \label{fig:cluster}
\end{figure*}

For an ideal stationary periodic signal, the feature set is
time-invariant and remains constant since it has been projected to the
frequency domain.  Even in a practical setting where some noise is
inevitable, the feature set does not vary much if the overall
evolution remains constant.  In other words, the feature set in each
measurement should only fluctuate around its mean and at a very small
scale.  Fig.~\ref{fig:pca:2} plots the variation of system-level
features with the mean subtracted from the values.  The lower figure
shows that the feature distribution during normal operations is very
narrow and Gaussian-like.  In the upper figure, we can see all the
corresponding blue points scattered around the origin within a very
small area.  The big red cross in the top-right corner marks the
deviant feature measured when a Sybil attack was launch in September
2011~\cite{WangL:BitTorrentSecurity}.  We can see it deviates very far
away from the center and this deviation can be used as an indicator of
an anomaly.  Note that this only indicates a possible anomaly but does
not (necessarily) yield information about the nature of the anomaly.

Therefore, we can take advantage of the stability of the
frequency-based feature and use it as an anomaly indicator.
Technically, it is done by reusing the similarity metric introduced in
Section~\ref{sec:core:feature} against a signal itself, namely
$S_{a,\bar{a}}$.

\begin{mydef}
  Anomaly indicator for an entity $a$ is defined as $S_{a,\bar{a}}
  = \parallel F_a - \mu_a \parallel $, where $F_a$ and $\mu_{a}$
  denote the current measurement on $F^*_a$ and its average value
  respectively.
\end{mydef}

By definition $S_{a,\bar{a}}$ is the deviation of the current measured
$F^*_a$ to its own average.  In other words, it measures how similar a
signal to itself on average in each measurement.  In practice, sample
mean is used in place of the true average.  An alarm can be triggered
if the deviation is large enough, e.g., $S_{a,\bar{a}} > 3 \sigma_a$
and $\sigma^2_a$ denotes the variance of $F^*_a$.  In the actual
implementation, we incorporate supervised machine learning technique
to achieve good balance between sensitivity and fall-out rate.

\section{Evaluation in the Wild}
\label{sec:exp}

In the following, we briefly report the performance of the proposed
method on a realistic monitoring system with some selected results.
We evaluate our method in terms of its effectiveness of clustering
similar user behaviors and the accuracy of capturing anomalies.  We
extended the basic system proposed in~\cite{6688697} by incorporating
the feature extraction and anomaly detection modules, and
reimplemented it on Spark.  The system is not only able to quickly
respond to the continuous real-time samples, but also capable of fast
processing terabytes historical archives.

\subsection{Discover Similarity in Use Patterns}
\label{sec:exp:cluster}

The evolution of network size represents how and when the system is
accessed and used, which further reflects the characteristics of the
use patterns from a group of geographically-close users.  The
frequency-based feature is expected to capture the similarity in such
patterns and reflect in the proposed similarity metric defined in
Section~\ref{sec:core:feature}.

To gain an intuitive understanding of the effectiveness of such
mechanisms, we use the first two principal components in the feature
set and plot them in Fig.\ref{fig:cluster:1}.  Countries are marked
with different colors according to their geographical regions, and
their coordinates in the figure are determined by the two principal
components.  At the first glance, most of the countries gather
together roughly based on their geographical proximity.  This is
understandable since the geographical closeness usually indicates that
the countries share social, cultural, and economic bonds which lead to
similar user behavior.  For example, among the European countries,
most East European countries stay at the right of the figure while the
most Western ones gather in the middle.

However, we can see some interesting phenomena among some sets of
countries.  Comparing Japan, Taiwan, and Korea, we see that Korea is
mapped to a very different place in Figure~\ref{fig:cluster:1} and if
we compare the actual daily patterns of these three countries, as
shown in Figure~\ref{fig:cluster:2}, we see that the Korean pattern is
clearly different from that of Japan or Taiwan.  Similar observation
holds when comparing Korea with Singapore or Hong Kong which map close
to Japan and Taiwan in Figure~\ref{fig:cluster:1}.  Likewise,
Figure~\ref{fig:cluster:3} compares USA, Canada, and Russia, and the
visual differences in the patterns are also reflected in the placement
of the three countries in Figure~\ref{fig:cluster:1}.  (USA and Canada
are towards the left, Russia is in the middle.)

We have compared other groupings of countries and have seen that
countries that map close to each other in Figure~\ref{fig:cluster:1}
have visually similar daily patterns.  However, data in
Figure~\ref{fig:cluster:1} only serves to classify countries into
different groupings.  The distances between different countries do not
really serve as an indicator of \emph{how} different their daily
patterns are; instead it only indicates the existence of a difference.



\subsection{Detecting Nontrivial Anomalies with Learning}
\label{sec:exp:anomaly}

\begin{figure*}[!tb]
  \centering \subfloat[ROC space of anomaly
  detection.]{\label{fig:ab:1}\includegraphics[width=5.2cm]{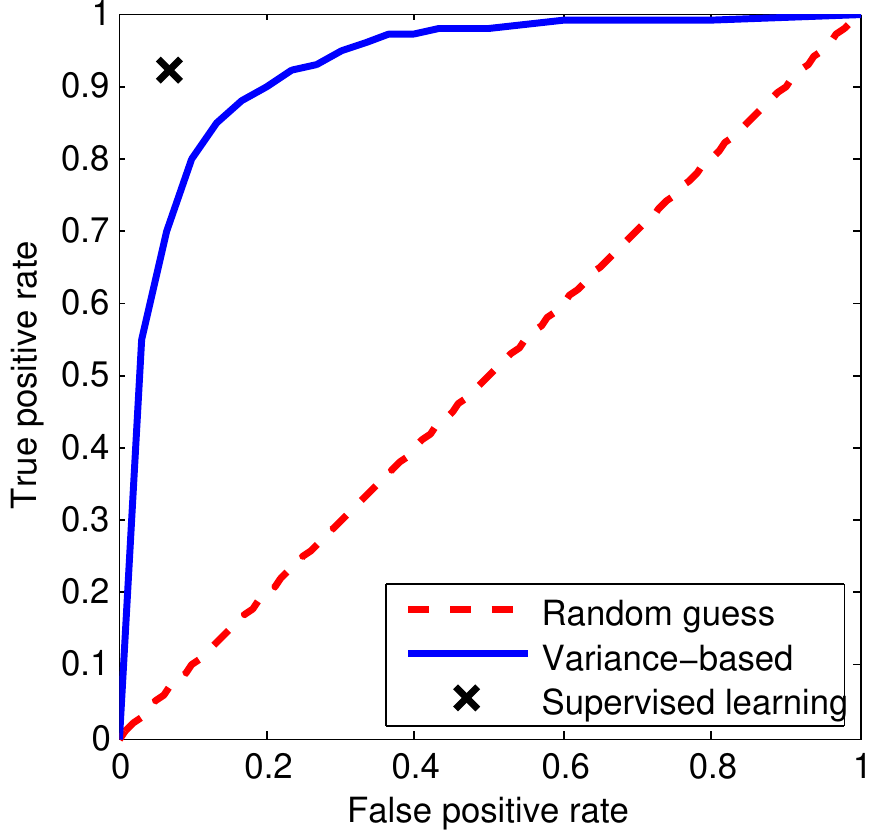}}
  \quad \subfloat[T\=ohoku earthquake in Japan
  2011.]{\label{fig:ab:2}\includegraphics[width=6.2cm]{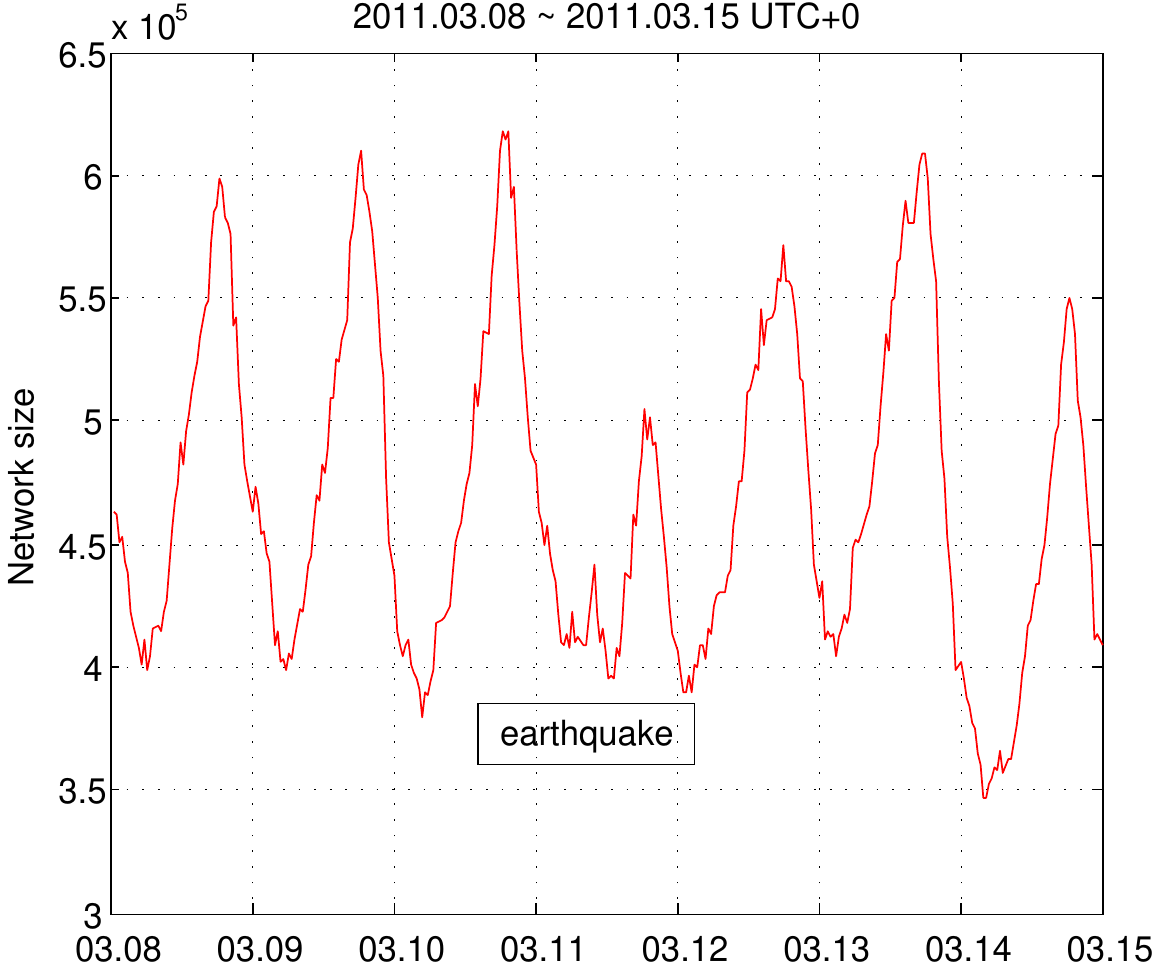}}
  \quad \subfloat[Behavior drift in 2013
  Christmas.]{\label{fig:ab:3}\includegraphics[width=11.5cm]{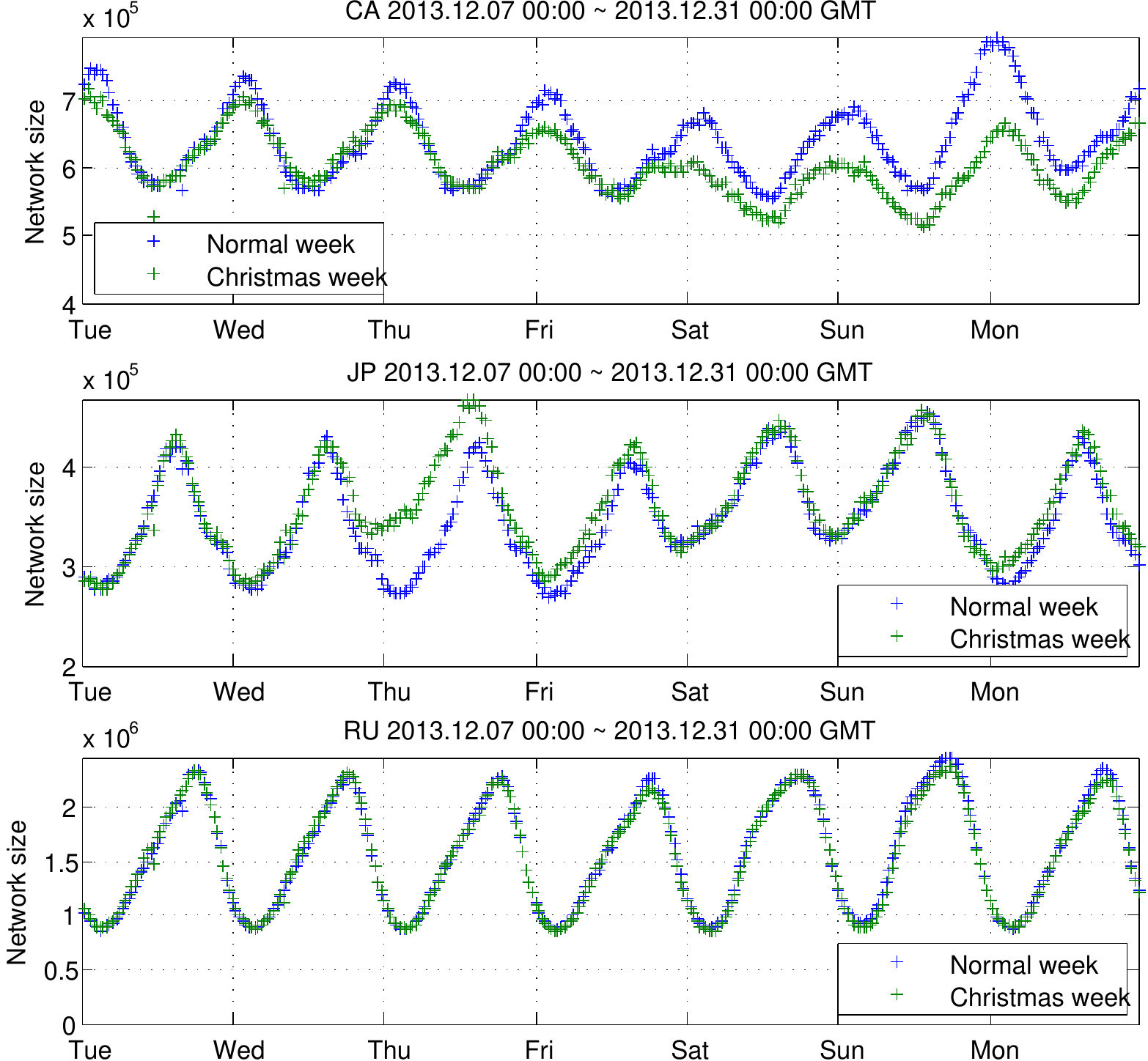}}
  \caption{Accuracy of anomaly detection with different threshold
    tuning techniques, and some captured real-world events.}
  \label{fig:ab}
\end{figure*}

Despite different potential causes, the system anomalies are reflected
as statistically significant changes in the feature by definition.
Anomaly detection takes advantage of the stability of frequency-based
features, and issues an alarm when the deviation is large enough.  As
we have introduced in Section~\ref{sec:core:anomaly}, the threshold of
triggering an alarm can either be manually set or automatically learnt
from labelled data.  The blue solid line in the Fig.~\ref{fig:ab:1}
shows the performance by simply varying variance-based threshold in
the ROC space.  Recall that ROC curve shows the trade-off between
sensitivity (true positive rate) and fall-out (false positive rate) in
a classification system and increasing the threshold improves fall-out
by sacrificing sensitivity.  The red dashed line in
Fig.~\ref{fig:ab:1} represents a pure guess, which serves as a
baseline that any rational algorithm should be above and the top-left
corner represents perfect accuracy.  Our evaluation shows that simple
variance-based threshold can already achieve good accuracy.  In the
actual system, we adopted an naive Bayesian classifier trained on a
data set of 500 labelled anomalies, and its accuracy is plotted as a
black cross in the same Fig.~\ref{fig:ab:1}.  As we can see,
frequency-based feature with basic supervised learning leads to a
high-accuracy detection module.

Besides the large-scale Sybil attack mentioned in
Section~\ref{sec:core:anomaly} and many others, our anomaly detection
system also successfully captured a lot of interesting real-world
events.  For example, there was a severe earthquake happened in
T\=ohoku, Japan on March 11th, 2011.  Figure~\ref{fig:ab:2} shows the
network evolution of the corresponding week.  Our feature extraction
method would have indicated an alarm due to the large-scale network
breakdown which caused tens of thousands of computers disconnected
from the network.  While Figure~\ref{fig:ab:2} illustrates the impacts
from natural causes, Figure~\ref{fig:ab:3} on the right shows how the
social and cultural events can change user behaviors.
Figure~\ref{fig:ab:3} plots the Christmas week in 2013 with green
lines for Canada, Japan, and Russia (from top to bottom).  To
highlight the pattern drift, the network evolution of a normal week is
also plotted in the same figure with blue line.  Interestingly, we
notice while there was significant drop in the network size during
Christmas in Canada (actually in most western countries), this number
increased quite a lot in Japan (also in many other Asian countries).
Russia, on the other hand, did not change much during this time
period.  One possible explanation could be that in Russia Christmas is
celebrated according to the Orthodox calendar, meaning that it falls a
few weeks later than in Western countries; in other words, the week
shown in Figure~\ref{fig:ab:3} is simply a normal week compared with
another normal week.

\subsection{Summary}
\label{sec:summary}

As the results above show, the features extracted via Fourier
transform allow us to look at the data from another point of view and
observe different phenomena more easily.  For example, characterizing
the diurnal patterns in different countries is readily visible in
Figure~\ref{fig:cluster} without the need to compare every country
pairs visually.

\section{Conclusion}
\label{sec:conclusion}

In this paper, we showed that Fourier transform can be used as a
simple yet powerful tool to extract representative information from
time series data in P2P measurements.  Applying Fourier transform on
network size evolution data, we showed that this seemingly trivial
statistic contains rich information.  The extracted frequency-based
feature can be effectively used to characterize user behaviors and
detect system anomalies.  We implemented a monitoring system with the
proposed algorithm and evaluated its actual performance in the
realistic environment with many interesting real-world findings.  Our
algorithm successfully discovered nontrivial use patterns and captured
anomalies with high accuracy.  In the future, we plan to apply this
technique on a broader range of data sets and also explore its other
potential applications in P2P monitor and measurement research.


\section*{Appendix}
\footnotesize
\begin{table}[!hb]
  \centering
  \begin{tabular}{|l|l|l|l|l|l|l|l|l|l|l|l|l|l|l|l|l|l|l|l|}
    \hline
    & BR & CA & CN & EG & FI & GB & HK & IL & IN & IT & JP & KR & NL & PL & RU & SE & US \\
    \hline
BR & 0.00 & 8.92 & 6.00 & 5.56 & 4.23 & 5.16 & 6.46 & 9.10 & 4.46 & 1.81 & 7.34 & 3.61 & 3.02 & 1.09 & 1.69 & 3.95 & 9.31 \\
CA & 8.92 & 0.00 & 2.95 & 3.39 & 4.69 & 3.77 & 2.59 & 0.19 & 4.48 & 7.16 & 1.72 & 5.78 & 5.92 & 8.22 & 7.33 & 4.97 & 0.39 \\
CN & 6.00 & 2.95 & 0.00 & 0.44 & 1.78 & 0.85 & 1.13 & 3.13 & 1.54 & 4.28 & 1.64 & 2.88 & 3.06 & 5.37 & 4.47 & 2.09 & 3.34 \\
EG & 5.56 & 3.39 & 0.44 & 0.00 & 1.35 & 0.42 & 1.35 & 3.56 & 1.10 & 3.84 & 2.00 & 2.48 & 2.63 & 4.94 & 4.03 & 1.67 & 3.78 \\
FI & 4.23 & 4.69 & 1.78 & 1.35 & 0.00 & 0.94 & 2.31 & 4.87 & 0.31 & 2.49 & 3.16 & 1.65 & 1.28 & 3.59 & 2.68 & 0.34 & 5.08 \\
GB & 5.16 & 3.77 & 0.85 & 0.42 & 0.94 & 0.00 & 1.53 & 3.94 & 0.71 & 3.43 & 2.30 & 2.22 & 2.21 & 4.52 & 3.62 & 1.25 & 4.15 \\
HK & 6.46 & 2.59 & 1.13 & 1.35 & 2.31 & 1.53 & 0.00 & 2.79 & 2.19 & 4.66 & 0.90 & 3.75 & 3.44 & 5.68 & 4.82 & 2.53 & 2.97 \\
IL & 9.10 & 0.19 & 3.13 & 3.56 & 4.87 & 3.94 & 2.79 & 0.00 & 4.66 & 7.35 & 1.91 & 5.94 & 6.11 & 8.41 & 7.52 & 5.16 & 0.22 \\
IN & 4.46 & 4.48 & 1.54 & 1.10 & 0.31 & 0.71 & 2.19 & 4.66 & 0.00 & 2.75 & 3.00 & 1.62 & 1.56 & 3.86 & 2.95 & 0.65 & 4.87 \\
IT & 1.81 & 7.16 & 4.28 & 3.84 & 2.49 & 3.43 & 4.66 & 7.35 & 2.75 & 0.00 & 5.55 & 2.36 & 1.24 & 1.12 & 0.22 & 2.19 & 7.55 \\
JP & 7.34 & 1.72 & 1.64 & 2.00 & 3.16 & 2.30 & 0.90 & 1.91 & 3.00 & 5.55 & 0.00 & 4.48 & 4.32 & 6.58 & 5.71 & 3.40 & 2.08 \\
KR & 3.61 & 5.78 & 2.88 & 2.48 & 1.65 & 2.22 & 3.75 & 5.94 & 1.62 & 2.36 & 4.48 & 0.00 & 1.77 & 3.40 & 2.58 & 1.68 & 6.16 \\
NL & 3.02 & 5.92 & 3.06 & 2.63 & 1.28 & 2.21 & 3.44 & 6.11 & 1.56 & 1.24 & 4.32 & 1.77 & 0.00 & 2.31 & 1.41 & 0.96 & 6.31 \\
PL & 1.09 & 8.22 & 5.37 & 4.94 & 3.59 & 4.52 & 5.68 & 8.41 & 3.86 & 1.12 & 6.58 & 3.40 & 2.31 & 0.00 & 0.91 & 3.27 & 8.60 \\
RU & 1.69 & 7.33 & 4.47 & 4.03 & 2.68 & 3.62 & 4.82 & 7.52 & 2.95 & 0.22 & 5.71 & 2.58 & 1.41 & 0.91 & 0.00 & 2.37 & 7.72 \\
SE & 3.95 & 4.97 & 2.09 & 1.67 & 0.34 & 1.25 & 2.53 & 5.16 & 0.65 & 2.19 & 3.40 & 1.68 & 0.96 & 3.27 & 2.37 & 0.00 & 5.36 \\
US & 9.31 & 0.39 & 3.34 & 3.78 & 5.08 & 4.15 & 2.97 & 0.22 & 4.87 & 7.55 & 2.08 & 6.16 & 6.31 & 8.60 & 7.72 & 5.36 & 0.00 \\
    \hline
  \end{tabular}
  \caption{The number in the table is a numerical presentation of Fig.\ref{fig:cluster:1}, which shows the Euclidean distance of any two selected countries. Small distance indicates there is a high degree of similarity between the two countries. }
  \label{tab:lvl:sim}
\end{table}

\end{document}